# The $^{30}$Si Mole Fraction of a Silicon Material Highly Enriched in $^{28}$Si Determined by Instrumental Neutron Activation Analysis

Giancarlo D'Agostino[1]*, Marco Di Luzio[1-2], Giovanni Mana[1], Massimo Oddone[2], Axel Pramann[3], and Michele Prata[4]

1 Istituto Nazionale di Ricerca Metrologica (INRIM), Strada delle Cacce 91, 10135 Torino, Italy

2 Department of Chemistry, University of Pavia, via Taramelli 12, 27100 Pavia, Italy

3 Physikalisch-Technische Bundesanstalt (PTB), Bundesallee 100, 38116 Braunschweig, Germany

4 Laboratorio Energia Nucleare Applicata (LENA), University of Pavia, via Aselli 41, 27100 Pavia, Italy

**ABSTRACT:** The latest determination of the Avogadro constant, carried out by counting the atoms in a pure silicon crystal highly enriched in $^{28}$Si, reached the target $2 \times 10^{-8}$ relative uncertainty required for the redefinition of the kilogram based on the Planck constant. The knowledge of the isotopic composition of the enriched silicon material is central; it is measured by isotope dilution mass spectrometry. In this work, an independent estimate of the $^{30}$Si mole fraction was obtained by applying a relative measurement protocol based on Instrumental Neutron Activation Analysis. The amount of $^{30}$Si isotope was determined by counting the 1266.1 keV $\gamma$-photons emitted during the radioactive decay of the radioisotope $^{31}$Si produced via the neutron capture reaction $^{30}$Si(n,$\gamma$)$^{31}$Si. The $x(^{30}\text{Si}) = 1.043(19) \times 10^{-6}$ mol mol$^{-1}$ is consistent with the value currently adopted by the International Avogadro Coordination.

Since 1889, the unit of mass of the International System of Units (SI) is a material artifact called the international prototype of the kilogram (IPK). While a unit of measurement should be invariant, it is suspected a relative mass drift of about $5 \times 10^{-8}$ over a century between the IPK and its national copies.[1] Given the paramount importance of the kilogram within the SI, an instability of the IPK mass is widely recognized as unacceptable.

The existing issue would be solved by a redefinition of the kilogram in terms of a fundamental constant of nature. With the purpose of assuring continuity to mass metrology, it has been agreed that the relative uncertainty of any new realization of the unit must not exceed $2 \times 10^{-8}$. In this framework, there is an ongoing attempt to redefine the kilogram in terms of an exact, fixed value of the Planck constant, $h$.[2-4]

At present, there are two independent techniques capable to determine $h$ to within the required accuracy. The first is based on the watt balance (WB) method[5,6] and the latter in based on the x-ray crystal density (XRCD) method.[7,8] The WB approach determines $h$ by measuring the electromagnetic force necessary to support 1 kg mass whereas the XRCD approach determines $h$ by measuring the Avogadro constant, $N_A$, which is linked to $h$ through the accurately known molar Planck constant, $N_A h$.

Recently, the XRCD experiment was carried out with upgraded measurement protocols and apparatuses using two 1 kg spheres made from a highly pure silicon single crystal, called AVO28, isotopically enriched in $^{28}$Si.[9] The Avogadro constant was determined with a relative uncertainty of $1.9 \times 10^{-8}$ by measuring the mass, $m$, the volume, $V$, the lattice parameter, $a$, and the molar mass, $M$, of each sphere. The correspondent $h$ value achieved the target $2 \times 10^{-8}$ relative uncertainty established by the Consultative Committee for Mass and Related Quantities (CCM) for the new definition of the kilogram.[10]

The molar mass was obtained from a number of measurements of the isotopic composition of the silicon material, i.e. the mole fractions $x(^{28}\text{Si})$, $x(^{29}\text{Si})$ and $x(^{30}\text{Si})$, performed by the Physikalisch-Technische Bundesanstalt (PTB), the National Metrology Institute of Japan (NMIJ) and the National Institute of Standards and Technology (NIST). The data were collected with multi-collector inductively coupled plasma mass spectrometers (MC ICP-MS) using the 'virtual element' isotope dilution mass spectrometry (VE-IDMS) technique.[11-13]

The adopted measurement protocol is the outcome of a significant effort dedicated to the development of the method used for the determination of the molar mass. The comparison of the results obtained in several experiments carried out in different laboratories[14-18] showed a general agreement with the only exception of a significant discrepancy observed by the Institute for National Measurement Standards, National Research Council Canada (NRC).

A contamination by natural silicon of the dissolved samples producing the consistent results was formerly argued to explain the inconsistency of the NRC datum.[15,19] However, a subsequent careful analysis of the analytical method suggested that the use of alkaline solutions containing sodium for sample dissolution and dilution affects the result.[20] Accordingly, high levels of aqueous sodium hydroxide (NaOH) must be avoided or, in alternative, tetramethylammonium hydroxide (TMAH) should be used instead of NaOH.[18]

Given that (i) all the available molar mass data have been obtained with the VE-IDMS technique and that (ii) the possible contamination of the $^{28}$Si-enriched samples by natural silicon is still under consideration, the application of an in-solid analytical technique based on a different method would be helpful to verify the available data and to solve the existing disagreement.

Accordingly, the use of neutron activation to determine the $x(^{30}$Si$)$ of a silicon material isotopically enriched in $^{28}$Si has been investigated. The experimental data collected with a natural silicon sample indicated that the measurement might reach a percent level relative uncertainty[21], corresponding to a contribution at $10^{-9}$ level to the relative uncertainty of $M$.

Since the anticipated relative uncertainty of the $x(^{30}$Si$)$ measurement is comparable to the relative uncertainties of the $x(^{30}$Si$)$ data collected by the VE-IDMS technique, the Istituto Nazionale di Ricerca Metrologica (INRIM) carried out a neutron activation analysis to determine the $x(^{30}$Si$)$ value of a sample of the silicon material isotopically enriched in $^{28}$Si and used to manufacture the 1 kg spheres. This paper gives the details of the experiment and the result of the measurement.

**Determination of the $^{30}$Si Mole Fraction.** The neutron activation analysis is based on the conversion of stable nuclei to radioactive nuclei via nuclear reactions and the detection of the $\gamma$-photons emitted during the decay of the reaction products.[22] The use of the non-destructive approach, called instrumental neutron activation analysis (INAA), avoids the dissolution of the sample and decreases the risk of contamination compared to other techniques in which the sample has to be dissolved.

The application of INAA for the determination of the $^{30}$Si mole fraction is feasible by counting the 1266.1 keV $\gamma$-photons emitted during the radioactive decay of $^{31}$Si produced by activation of $^{30}$Si via neutron capture reaction $^{30}$Si$(n,\gamma)^{31}$Si. The proof of concept and estimation of the achievable uncertainty in case of pure silicon materials highly enriched in $^{28}$Si were reported.[21]

The measurement protocol consists in the simultaneous irradiation of the $^{28}$Si-enriched silicon sample and a natural silicon sample acting as a calibrator for the $^{30}$Si isotope. The effects due to the neutron self-shielding during the neutron irradiation, the emission self-absorption during the $\gamma$-counting are considerably reduced by using a calibration sample having the same matrix, i.e. a silicon single crystal, shape and dimensions of the $^{28}$Si-enriched sample.

The $x(^{30}$Si$)$ value is determined using the following model function:

$$x(^{30}\text{Si}_{28\text{Si}}) = \kappa_{td} \, \kappa_R \, \kappa_\varepsilon \, \kappa_{ss} \, \kappa_{sa} \, \kappa_g \, \frac{C_{28\text{Si}}(t_{d\,28\text{Si}})}{C_{nat\text{Si}}(t_{d\,nat\text{Si}})} x(^{30}\text{Si}_{nat\text{Si}}) \\ \times \frac{m_{nat\text{Si}}}{m_{28\text{Si}}} \frac{M_{28\text{Si}}}{M_{nat\text{Si}}} \quad (1)$$

where the subscripts 28Si and natSi refer to the $^{28}$Si-enriched and natural silicon sample, respectively, $C(t_d)$ is the 1266.1 keV full-energy $\gamma$-peak detection count rate at a time $t_d$ after the end of the irradiation, $m$ is the mass of the silicon sample and $M$ is the molar mass. The correction factors $\kappa_{td} = e^{-\lambda(t_{d\,natSi}-t_{d\,28Si})}$, $\kappa_R = R_{natSi}/R_{28Si}$, $\kappa_\varepsilon = \varepsilon_{natSi}/\varepsilon_{28Si}$, $\kappa_{ss} = k_{ss\,natSi}/k_{ss\,28Si}$, $\kappa_{sa} = k_{sa\,natSi}/k_{sa\,28Si}$ and $\kappa_g = k_{g\,natSi}/k_{g\,28Si}$ take the differences of decay time, reaction rate, detection efficiency, self-shielding, self-absorption and geometry of the silicon samples into account.

In details, $\lambda$ is the decay constant of $^{31}$Si, $R$ is the $(n,\gamma)$ reaction rate per target $^{30}$Si atom, $\varepsilon$ is the detection full-energy $\gamma$ efficiency for a point-like source located at the center of mass of the sample, $k_{ss}$, $k_{sa}$, and $k_g$ are the irradiation self-shielding, the emission self-absorption and the geometry factors, respectively.

It is noteworthy that the second term of equation (1) depends on the isotopic composition of the $^{28}$Si-enriched sample through

$$M_{28\text{Si}} = M(^{28}\text{Si}) + [M(^{29}\text{Si}) - M(^{28}\text{Si})] x(^{29}\text{Si}_{28\text{Si}}) \\ + [M(^{30}\text{Si}) - M(^{28}\text{Si})] x(^{30}\text{Si}_{28\text{Si}}) \quad (2)$$

where $M(^{28}$Si$)$, $M(^{28}$Si$)$ and $M(^{28}$Si$)$ are the accurately known molar masses of the silicon isotopes. Accordingly, equation (1) sets only a constrain between the $x(^{29}$Si$_{28Si})$ and $x(^{30}$Si$_{28Si})$ values. However, in case of a material highly enriched in $^{28}$Si, we can neglect the dependence of $M_{28Si}$ on the $x(^{29}$Si$_{28Si})$ and $x(^{30}$Si$_{28Si})$ values.

**Detection Count Rate.** The counting of the $\gamma$-photons emitted by $^{31}$Si is performed using germanium detectors. The count rate $C(t_d)$ at a time $t_d$ after the end of the irradiation is obtained by averaging $n$ values, $C_i(t_d)$, acquired in a $\gamma$-spectrometry sequence starting at $t_d$ and consisting of $n$ consecutive counts performed during the decay of $^{31}$Si. More expressly, each $i^{th}$ count rate value, $C_i(t_d)$, extrapolated at $t_d$ from the $i^{th}$ count of the sequence, starting at $t_d$ and lasting $t_{c\,i}$, is

$$C_i(t_d) = \frac{\lambda \, n_{c\,i}}{e^{-\lambda(t_{d\,i}-t_d)}(1-e^{-\lambda t_{c\,i}})} \frac{t_{c\,i}}{t_{c\,i}-t_{dead\,i}} \quad (3)$$

where the subscripts natSi and 28Si have been omitted, $\lambda$ is the $^{31}$Si decay constant, $n_{c\,i}$ and $t_{dead\,i}$ are the net count of the 1266.1 keV $\gamma$-peak and the detection dead time of the $i^{th}$ count, respectively.

**Neutron Activation Rate.** The spatial variation of the neutron energy spectrum in a reactor depends on the position of fuel elements and control rods. Typically, the integral (total) flux reaches a maximum at the center of the core. Since the reaction rate per target nucleus, $R$, depends on the neutron energy spectrum irradiating the samples, it is essential to monitor the non-uniformity of the neutron flux at the irradiation positions.

The simultaneous irradiation of the $^{28}$Si-enriched and natural silicon samples makes only the variation of amplitude of the neutron energy spectrum relevant. Since the reaction $^{30}$Si$(n,\gamma)^{31}$Si has a cross section with a $E^{-1/2}$ energy dependence, the reaction rate can be expressed according to the Høgdahl convention[23], $R = \Phi_{th}(\sigma_0 + I_0(\alpha)/f)$, where $\Phi_{th}$ is the thermal-neutron flux irradiating the sample, $\sigma_0$ is the $(n,\gamma)$ cross section at 0.0253 eV, $I_0(\alpha)$ is the resonance integral for a $1/E^{1+\alpha}$ spectrum and $f$ is the thermal to epithermal neutron flux ratio. Consequently, the $\kappa_R$ correction factor can be estimated by measuring the ratio of the thermal fluxes at the irradiation positions.

To this purpose, flux monitors consisting of a known mass of an isotope having a $(n,\gamma)$ cross section with a $E^{-1/2}$ energy dependence are located near the samples during the irradiation. The following activation equation applies:

$$\kappa_R = \frac{\Phi_{\text{th natSi}}}{\Phi_{\text{th 28Si}}} = (\kappa_{\text{m-td}}\ \kappa_{\text{m-}\varepsilon}\ \kappa_{\text{m-ss}}\ \kappa_{\text{m-sa}}\ \kappa_{\text{m-g}})^{-1}$$
$$\times \frac{C_{\text{m-natSi}}(t_{\text{d m-natSi}})}{C_{\text{m-28Si}}(t_{\text{d m-28Si}})}\frac{m_{\text{m-28Si}}}{m_{\text{m-natSi}}} \quad (4)$$

where the subscripts m-28Si and m-natSi refer to the monitors located near the $^{28}$Si-enriched and natural silicon samples, respectively.

## EXPERIMENTAL SECTION

**Materials.** A cylindrical sample of the $^{28}$Si-enriched AVO28 crystal produced within the International Avogadro Coordination (IAC) project was cut by the PTB. The sample identification is Si28-10-Pr11 part 6.9.3. The calibration material was provided by INRIM and cut by the PTB. It consisted of a cylindrical sample of the WASO04 natural silicon crystal (sample identification WASO04 96S92 part 2) previously used for the determination of the Avogadro constant. The chemical properties of the WASO04 material have been widely studied in the framework of the determination of the Avogadro constant. In particular, the isotopic composition and the purity have been newly measured by PTB using multicollector-ICP-mass spectrometry and investigated by INRIM using INAA[24], respectively. Accordingly, this material is suitable as a calibrator for the $^{30}$Si isotope. After cutting, the diameter, $\phi$, and length, $l$, of the samples were $\phi_{\text{AVO28}} = 9.586$ mm, $\phi_{\text{WASO04}} = 9.588$ mm and $l_{\text{AVO28}} = 31.15$ mm, $l_{\text{WASO04}} = 31.20$ mm, where the subscripts AVO28 and WASO04 refers to the AVO28 and WASO04 material. Two samples, about 31 mm length, of a high-purity Co-Al wire (Reactor Experiment, code n. 604D, 0.46% Co mass fraction, 99.9313% purity, 0.381 mm diameter) were cut to monitor the neutron flux at the irradiation positions.

**Preparation of the samples.** Both the silicon samples were sequentially (i) washed with trichloroethylene, acetone and deionized water, (ii) etched for 10 min with a solution 10:1 of nitric acid (assay 67-69%) and hydrofluoric acid (assay 47-51%), and finally (iii) rinsed in deionized water, ethylalcohol and acetone to eliminate the surface contamination. This sequence was iterated by monitoring the mass variation of the sample until the removed silicon material was about 200 mg, corresponding to about 80 μm external thickness. The final masses of the silicon samples, $m_{\text{AVO28}} = 4.99060(3)$ g and $m_{\text{WASO04}} = 5.01260(3)$ g, were measured using a digital analytical balance calibrated with SI-traceable weights. Here and hereafter, unless otherwise specified, the brackets refer to the standard uncertainty. The AVO28 sample was handled before the WASO04 sample and, after etching, they were kept physically separated to limit the $^{30}$Si contamination of the $^{28}$Si-enriched material.

The Co-Al monitors were sealed in two polyethylene microtubes. The masses of the monitors used for the AVO28 and WASO04 samples, 9.090(3) mg and 9.110(3) mg, respectively, were measured using a digital analytical balance calibrated with SI-traceable weights. The corresponding masses of the cobalt, i.e. the monitor element, are $m_{\text{m-AVO28}} = 40.905(13)$ μg and $m_{\text{m-WASO04}} = 40.995(13)$ μg.

The silicon samples were closed with their monitors in polyethylene irradiation vials (see figure 1a). The microtube surrounding the monitor avoids the contact between the silicon and monitor materials.

The vials, the microtubes, the tweezers used to handle the silicon and monitor materials, the perfluoroalkoxy alkanes beakers used to wash, etch and rinse the silicon materials were cleaned in an ultrasonic bath with 10% (mass fraction) $HNO_3$. All the chemicals were of ultrapure grade and the water was purified using a Millipore system ($\rho \geq 18$ MΩ).

**Neutron Irradiation.** The neutron irradiation lasted 6 hours and was performed in the 250 kW TRIGA Mark II reactor at the Laboratory of Applied Nuclear Energy (LENA) of the University of Pavia. The nominal thermal and epithermal neutron fluxes are about $6 \times 10^{12}$ cm$^{-2}$ s$^{-1}$ and $5.5 \times 10^{11}$ cm$^{-2}$ s$^{-1}$, respectively. The vials containing the silicon samples and the monitors were put in two polyethylene containers used for irradiation and located in the central thimble of the reactor. Figure 1b shows the position of the silicon samples with respect to the equator of the reactor core.

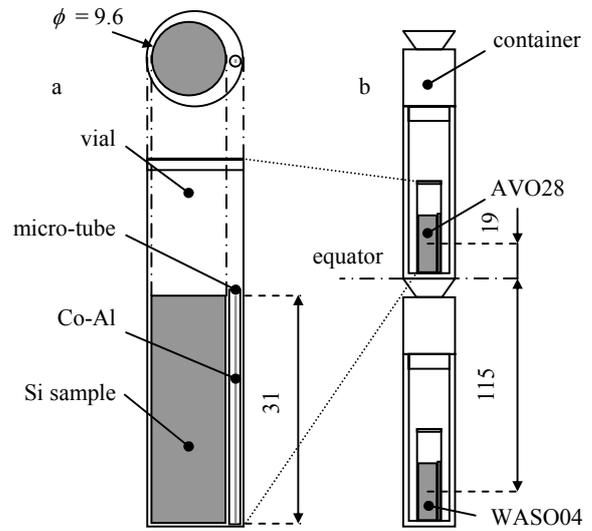

Figure 1. (a) Irradiation vial with the Si sample and the Co-Al monitor. (b) Position of the AVO28 and WASO04 samples with respect to the equator of the reactor core. Dimensions are in mm.

**Gamma Spectrometry.** After the neutron irradiation, the silicon samples and the monitors were extracted from the vials and rinsed with 10% (mass fraction) $HNO_3$. Four γ-spectrometry sequences were recorded with a coaxial HPGe detector ORTEC® GEM50P4-83 (66 mm crystal diameter, 50% relative efficiency, 1.90 keV FWHM resolution at 1332 keV), a digital signal processor ORTEC® DSPEC 502, and a personal computer running the software for data acquisition ORTEC® Gamma Vision.[25]

The first γ-spectrometry sequence concerned the 1332.5 keV $^{31}$Si γ-emission of the AVO28 sample and started at $t_{\text{d AVO28}} = 72.6$ min after the end of the irradiation. It consisted of 13 consecutive counts performed with a fixed counting window lasting 60 min. The second spectrometry sequence concerned the 1266.1 keV $^{31}$Si γ-emission of the WASO04 sample and started at $t_{\text{d WASO04}} = 1545.7$ min. It consisted of 36 counts performed by adjusting on-line the counting window to have 1% and 8% counting uncertainty for the first 29 and the latter 7 counts, respectively. The counting position of the silicon samples is displayed in figure 2a. The cylindrical sample is resting in a silt made in a PMMA holder fixed to a counting container. The container is centered on the detector and supported by three small spheres. The distance between the axis of the silicon sample and detector end-cup is 12 mm.

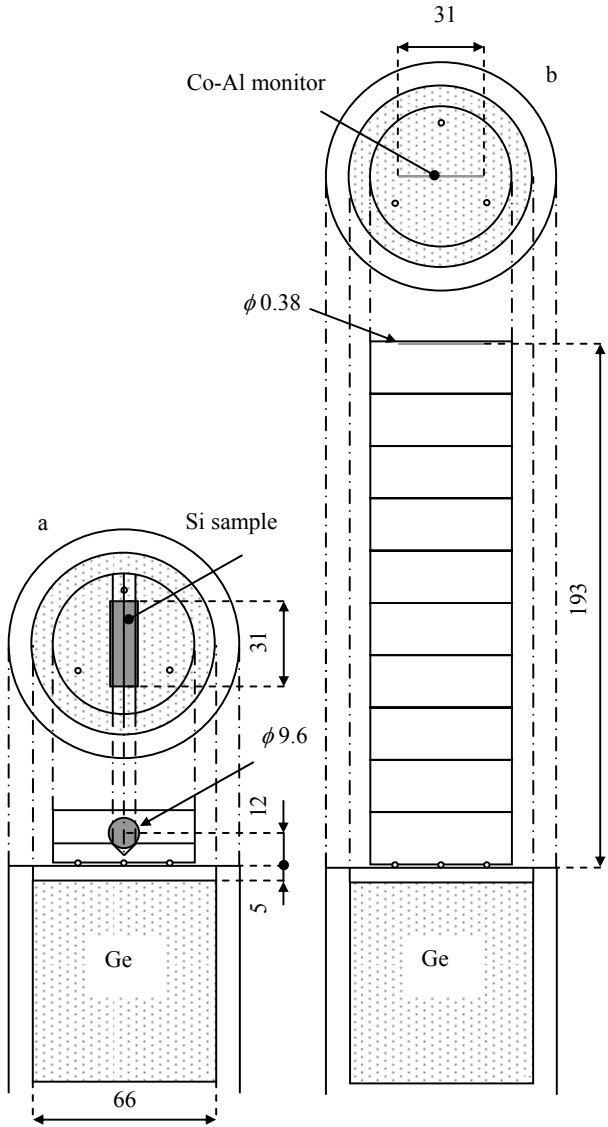

Figure 2. Position of the cylindrical Si sample (a) and of the Co-Al wire flux monitor (b) with respect to the Ge detector. Dimensions are in mm.

The third and fourth γ-spectrometry sequences concerned the $^{60}$Co γ-emission occurring at 1332.5 keV and produced by $^{59}$Co via (n,γ) reaction in the Al-Co monitors of the AVO28 and WASO04 samples, respectively. The sequences started at $t_{d\,m\text{-}AVO28}$ = 223.4 h and $t_{d\,m\text{-}WASO04}$ = 353.2 h after the end of the irradiation and consisted of 9 and 14 consecutive counts performed by adjusting on-line the counting window to have 0.2% counting uncertainty. The counting position of the monitors is displayed in figure 2b. The Co-Al monitor is fixed to the upper container of a stack of 10 containers centered on the detector and supported by three small spheres. The distance between the axis of the monitor and the detector end-cup is 193 mm.

The dead to counting time ratio, $t_{dead}/t_c$, of the detection system during the data collection was always below 10%, with the exception of the first 19 counts of the first sequence where it was between 26% and 10%. The stability of the detection system in terms of energy and resolution was checked before and after the γ-spectrometry measurements using a standard multi-gamma source located 5 cm from the detector end-cup.

## RESULTS AND DISCUSSION

**Gamma Peak Fitting.** The 1266.1 keV $^{31}$Si γ-peaks of the 49 spectra recorded during the first two sequences and the 1332.5 keV $^{60}$Co γ-peaks of the 23 spectra recorded during the latter two sequences were separately fitted using the following Gaussian function:

$$y = y_{0\,i} + \frac{n_{c\,i}}{\sigma\sqrt{\pi/2}} e^{-2\left(\frac{x-x_c}{\sigma}\right)^2} \qquad (5)$$

where $\sigma$ and $x_c$ are the standard deviation and centroid of the γ-peaks, $y_{0\,i}$ and $n_{c\,i}$ are the background and the net count of the γ-peak at the $i^{th}$ spectrum. The adjustment of the parameters was performed by sharing the $\sigma$ and $x_c$ values among the γ-peaks in the non-linear curve fitting algorithm implemented on the software OriginPro 7.5®.[26] The resulting $\sigma$ and $x_c$ values of the $^{31}$Si γ-peaks were 1.569 keV and 1266.51 keV, respectively, while the $\sigma$ and $x_c$ values of the $^{60}$Co γ-peaks were 1.615 keV and 1332.48 keV, respectively.

**Detection Count Rates.** The $i^{th}$ count rate, $C_i(t_d)$, was computed according to equation (3). The decay constants, $\lambda = \ln(2)/t_{1/2}$, were calculated using the half-life literature values, i.e. $t_{1/2}(^{31}\text{Si})$ = 157.36(26) min and $t_{1/2}(^{60}\text{Co})$ = 462.067(34) × 10$^2$ h.[27,28]

The resultant $^{31}$Si count rates recorded with the AVO28 sample and the WASO04 sample, extrapolated at $t_{d\,AVO28}$ and $t_{d\,WASO04}$, are reported in figure 3a and figure 3b, respectively. The error bars indicate the 95% confidence interval due to counting statistics.

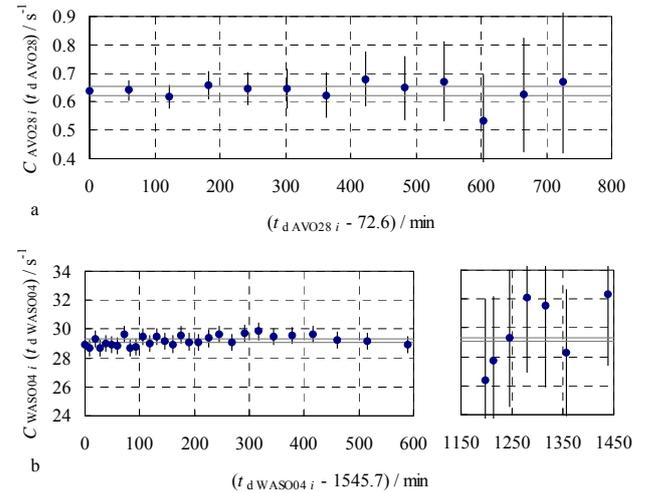

Figure 3. (a) $^{31}$Si count rates of the AVO28 sample extrapolated at $t_{d\,AVO28}$ = 72.6 min. (b) $^{31}$Si count rates of the WASO04 sample extrapolated at $t_{d\,WASO04}$ = 1545.7 min. The error bars indicate a 95% confidence interval. The horizontal lines show the 95% confidence interval associated to the weighted mean of the count rate values.

The weighted mean values of the $^{31}$Si count rates recorded with the AVO28 sample and with the WASO04 sample are $C_{AVO28}(t_{d\,AVO28}) = 0.6385(79)$ s$^{-1}$ and $C_{WASO04}(t_{d\,WASO04}) = 29.194(52)$ s$^{-1}$, respectively (uncertainties are due to counting statistics). The horizontal lines in figure 3 show the 95% confidence interval associated to the weighted mean of the $^{31}$Si count rates values. The agreement between the last 7 count rates recorded with the WASO04 sample (see figure 3b) and the weighted mean value proves that the fitting algorithm doesn't introduce a significant systematic error in case of reduced counting statistics.

The $^{60}$Co count rates recorded with the monitor of the AVO28 sample and with monitor of the WASO04 sample, extrapolated at $t_{d\,m\text{-}AVO28}$ and $t_{d\,m\text{-}WASO04}$ are reported in figure 4a and figure 4b, respectively. The error bars indicate the 95% confidence interval due to counting statistics.

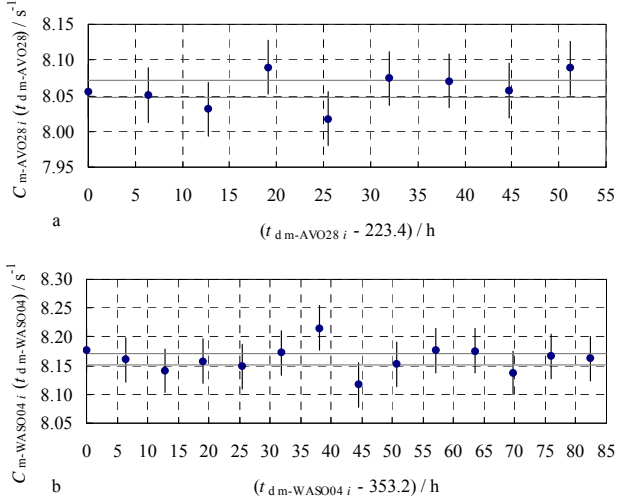

Figure 4. (a) $^{60}$Co count rates of the monitor of the AVO28 sample extrapolated at $t_{d\,m\text{-}AVO28} = 223.4$ h. (b) $^{60}$Co count rates of the monitor of the WASO04 sample extrapolated at $t_{d\,m\text{-}WASO04} = 353.2$ min. The error bars indicate a 95% confidence interval. The horizontal lines show the 95% confidence interval associated to the weighted mean of the count rate values.

The weighted mean values of the count rates recorded with the monitors of the AVO28 and WASO04 samples are $C_{m\text{-}AVO28}(t_{d\,m\text{-}AVO28}) = 8.0594(63)$ s$^{-1}$ and $C_{m\text{-}WASO04}(t_{d\,m\text{-}WASO04}) = 8.1612(51)$ s$^{-1}$, respectively (uncertainties are due to counting statistics). The horizontal lines in figure 4 indicate the 95% confidence interval associated to the weighted mean of the $^{60}$Co count rates values. It should be noted that the values obtained from the 7$^{th}$ and 8$^{th}$ count of the fourth $\gamma$-spectrometry sequence (figure 4b) are not in agreement with the weighed mean value. Since the analysis of the residuals of the corresponding fitted $\gamma$-peaks did not show a systematic anomaly and the detection system was properly working, the related data were not removed.

**Correction Factors.** The $\gamma$-spectrometry sequence of the WASO04 sample started 1473.1 min, i.e. about 10 times $t_{1/2}(^{31}$Si$)$, after the start of the $\gamma$-spectrometry sequence of the AVO28 sample. Accordingly, the $\kappa_{td}$ correction factor in equation (1) is 0.001521(16). The uncertainty is computed using the linear approximation of $\kappa_{td}$ and depends mainly on the uncertainty of the literature $t_{1/2}(^{31}$Si$)$ value.

After etching, the difference between the diameters and the lengths of the silicon samples is assumed lower than 0.05 mm and 0.1 mm, respectively. As a result, given that the linear attenuation coefficient of $\gamma$-photons in silicon at 1266.1 keV is about 0.1 cm$^{-1}$ and it is independent on the isotopic composition[29], the deviation of $\kappa_{sa}$ from the unit value is within $\pm 1 \times 10^{-4}$. The characterization of the Ge detector at the counting position showed that the deviation of $\kappa_g$ from the unit value is within $\pm 1.5 \times 10^{-4}$.

The following analytical expression, reported in literature[30], is adopted to calculate the (effective) neutron self-shielding factor

$$k_{ss} = \frac{f}{f + Q_0(\alpha)} k_{ss\,th} + \frac{Q_0(\alpha)}{f + Q_0(\alpha)} k_{ss\,ep}, \qquad (6)$$

where $Q_0(\alpha)$ is the ratio of the resonance integral to the thermal cross section of $^{30}$Si, $I_0(\alpha)/\sigma_0$, $k_{ss\,th}$ and $k_{ss\,ep}$ are the thermal and epithermal neutron self-shielding factors.

According to the cylindrical dimensions and isotopic composition of the silicon samples, $k_{ss\,th\,WASO04} = 0.988$, $k_{ss\,ep\,WASO04} = 0.995$, $k_{ss\,th\,AVO28} = 0.987$, $k_{ss\,ep\,AVO28} = 1.000$. Taking into account the characteristics of the neutron flux, from equation (6) it follows $k_{ss\,WASO04} = 0.9886$ and $k_{ss\,AVO28} = 0.9888$. As a result, we assume $\kappa_{ss} = 0.9998$ within $\pm 2 \times 10^{-4}$. It is worth mentioning that in our case, since $f$ is about one order higher than $Q_0(\alpha)$, the thermal and epithermal components of the effective neutron self-shielding factors are about 0.87 and 0.12, respectively

The counting container of the silicon sample is centered on the end-cup of the detector within 2 mm from the axis of the Ge crystal. The reproducibility of the vertical, horizontal and angular position is within 0.5 mm, 2 mm and 15°. The first of these is the main contributor to detection efficiency. Accordingly, $\kappa_\varepsilon = 1$ with a standard uncertainty of $6.6 \times 10^{-3}$.

Equation (4) was applied to the cobalt used to monitor the neutron flux. Taking into account the dimensional characteristics and the counting position of the monitors, $\kappa_{m\text{-}\varepsilon} = \kappa_{m\text{-}ss} = \kappa_{m\text{-}sa} = \kappa_{m\text{-}g} = 1$ with negligible uncertainties. Additionally, since $t_{d\,m\text{-}WASO04} - t_{d\,m\text{-}AVO28} = 129.8$ h, i.e. about $2 \times 10^{-3}$ times $t_{1/2}(^{60}$Co$)$, $\kappa_{m\text{-}td} = 0.998$ with negligible uncertainty. The corresponding $\kappa_R$ value is 1.0127(11). The uncertainty depends on count rates and masses of the monitors.

**Molar Masses of the Silicon Samples.** The $x(^{28}$Si$_{WASO04})$, $x(^{29}$Si$_{WASO04})$ and $x(^{30}$Si$_{WASO04})$ values are 0.92228(23) mol mol$^{-1}$, 0.04676(14) mol mol$^{-1}$ and 0.03096(11) mol mol$^{-1}$, respectively. According to the atomic masses of the silicon isotopes[31], the correspondent molar mass is $M_{WASO04} = 28.08549(34)$ g mol$^{-1}$.

The $x(^{29}$Si$_{AVO28})$ and $x(^{30}$Si$_{AVO28})$ values are expected to be about $4 \times 10^{-5}$ mol mol$^{-1}$ and $1 \times 10^{-6}$ mol mol$^{-1}$, respectively.[14,16-18] Therefore, in relation to equation (2), we assume $M_{AVO28} = M(^{28}$Si$) + 4.2 \times 10^{-5}$ g mol$^{-1}$ within $\pm 4.2 \times 10^{-5}$ g mol$^{-1}$.

**Estimate of the $^{30}$Si Mole Fraction.** From equation (1), it follows that the $^{30}$Si mole fraction of the AVO28 sample is $1.043(19) \times 10^{-6}$ mol mol$^{-1}$. The uncertainty budget calculated according to the Guide to the Expression of Uncertainty in Measurement[32] is reported in table 1.

**Table 1. Uncertainty budget of the $^{30}$Si mole fraction value of AVO28 material obtained by INAA. The input quantities $x_i$ are given in the text. The index column gives the relative contributions of $u(x_i)$ to the combined standard uncertainty, $u_c(y)$, of $x(^{30}Si_{AVO28})$.**

| Quantity | Unit | Value | Standard uncertainty | Index |
|---|---|---|---|---|
| $X_i$ | $[X_i]$ | $x_i$ | $u(x_i)$ | % |
| $\kappa_{td}$ | 1 | 0.001521 | 0.000016 | 34.9 |
| $\kappa_R$ | 1 | 1.0127 | 0.0011 | 0.4 |
| $\kappa_\varepsilon$ | 1 | 1.0000 | 0.0066 | 13.1 |
| $\kappa_{ss}$ | 1 | 0.99980 | 0.00012 | 0.0 |
| $\kappa_{sa}$ | 1 | 1.000000 | 0.000058 | 0.0 |
| $\kappa_g$ | 1 | 1.000000 | 0.000087 | 0.0 |
| $C_{AVO28}$ | s$^{-1}$ | 0.6385 | 0.0079 | 46.9 |
| $C_{WASO04}$ | s$^{-1}$ | 29.194 | 0.052 | 1.0 |
| $x(^{30}Si_{WASO04})$ | mol mol$^{-1}$ | 0.03096 | 0.00011 | 3.8 |
| $m_{WASO04}$ | g | 5.01260 | 0.00003 | 0.0 |
| $m_{AVO28}$ | g | 4.99060 | 0.00003 | 0.0 |
| $M_{AVO28}$ | g mol$^{-1}$ | 27.976969 | 0.000024 | 0.0 |
| $M_{WASO04}$ | g mol$^{-1}$ | 28.08549 | 0.00034 | 0.0 |
| $Y$ | $[Y]$ | $y$ | $u_c(y)$ | |
| $x(^{30}Si_{AVO28})$ | mol mol$^{-1}$ | 0.000001043 | 0.000000019 | 100.0 |

The main contributors to the overall uncertainty are (i) the $^{31}$Si count rate of the AVO28 sample, $C_{AVO28}(t_{d\ AVO28})$, (ii) the $\kappa_{td}$ correction factor and the (iii) $\kappa_\varepsilon$ correction factor.

**Comparative Results.** The published results concerning the isotopic composition of the AVO28 material collected with the VE-IDMS technique are summarized in table 2 together with the INAA result obtained in this study. The PTB$_{2011}$ and NRC$_{2012}$ data were achieved using NaOH for the preparation of the silicon solution. The most recent measurements, i.e. the PTB$_{2014}$, NMIJ$_{2014}$, NIST$_{2014}$ and PTB$_{2015}$ used TMAH instead of NaOH.

**Table 2. Isotopic composition determinations of the AVO28 material collected with VE-IDMS by different laboratories and the INRIM$_{2015}$ $x(^{30}Si)$ value obtained in this study with INAA. The references to the published paper are also given.**

| Lab. | $x(^{28}Si)$ (mol mol$^{-1}$) | $x(^{29}Si)$ (mmol mol$^{-1}$) | $x(^{30}Si)$ (mmol mol$^{-1}$) |
|---|---|---|---|
| PTB$_{2011}$[13] | 0.999 957 50(17) | 41.21(15) | 1.290(40) |
| NRC$_{2012}$[14] | 0.999 958 79(19) | 40.54(14) | 0.670(60) |
| PTB$_{2014}$[15] | 0.999 957 26(17) | 41.62(17) | 1.120(60) |
| NMIJ$_{2014}$[16] | 0.999 957 63(07) | 41.20(14) | 1.180(69) |
| NIST$_{2014}$[17] | 0.999 957 701(41) | 41.22 3(41) | 1.076(88) |
| PTB$_{2015}$[8] | 0.999 957 50(12) | 41.38(12) | 1.121(14) |
| INRIM$_{2015}$ | - | - | 1.043(19) |

The $x(^{29}Si)$ and $x(^{30}Si)$ values reported in table 2 are plotted in figure 5. The horizontal lines show the 95% confidence interval associated to the INRIM$_{2015}$ $x(^{30}Si)$ value. The discrepancy between the INRIM$_{2015}$ $x(^{30}Si)$ value and the NRC$_{2012}$ $x(^{30}Si)$ value is evident in figure 5a. Conversely, figure 5b points out that the INRIM$_{2015}$ $x(^{30}Si)$ value is consistent with the NIST$_{2014}$, the PTB$_{2014}$ values and to some extent with the NMIJ$_{2014}$ $x(^{30}Si)$ values.

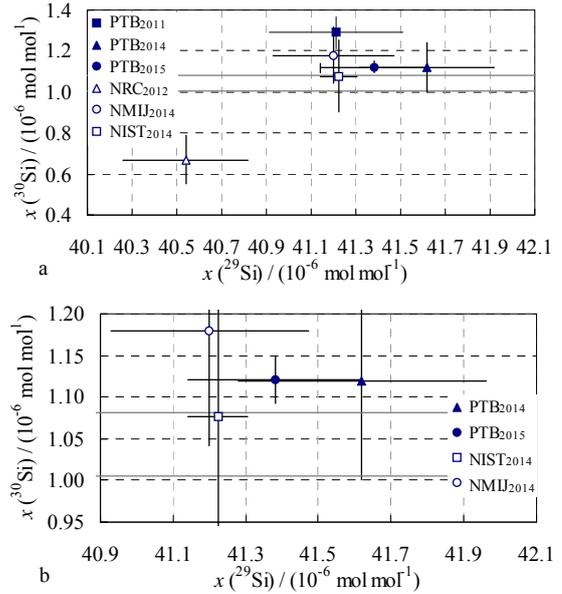

Figure 5. (a) The $x(^{29}Si)$ and $x(^{30}Si)$ values of the AVO28 material collected using the VE-IDMS technique. (b) A magnification of the PTB$_{2014}$, PTB$_{2015}$, NIST$_{2014}$ and NMIJ$_{2014}$ $x(^{29}Si)$ and $x(^{30}Si)$ values used to calculate the latest Avogadro constant. The error bars indicate a 95% confidence interval. The horizontal lines show the 95% confidence interval associated to the INRIM$_{2015}$ $x(^{30}Si)$ value.

## CONCLUSIONS

The INAA technique can be applied to determine the $^{30}$Si mole fraction of the silicon material highly enriched in $^{28}$Si used for the determination of the Avogadro constant (and, by implication, to the Planck constant). The immunity from natural silicon contamination of this in-solid analytical technique makes it a powerful test of the VE-IDMS molar-mass measurements.

The $x(^{30}Si)$ value obtained in this study reached a relative standard uncertainty lower than 2%, corresponding to a contribution to the relative uncertainty of the AVO28 molar mass of $1.3 \times 10^{-9}$, if adopted. Additionally, the INRIM$_{2015}$ $x(^{30}Si)$ value is consistent with the PTB$_{2014}$ and NIST$_{2014}$ $x(^{30}Si)$ values. This agreement rules out a significant contamination of the isotopic composition results collected by the VE-IDMS technique and used to calculate the molar mass of the AVO28 material adopted for the latest $N_A$ determination.[9]

The uncertainty budget pointed out that the count rate and the half-life of the radioactive $^{31}$Si atoms produced by the target $^{30}$Si atoms are the first two contributors to the measurement uncertainty. To enhance the $^{31}$Si count rate, we plan to repeat the experiment using the higher neutron flux of the 20 MW OPAL reactor at the Australian Nuclear Science and Technology Organisation (ANSTO). As well, the preliminary results we obtained by recording the decay of the $^{31}$Si showed that a more accurate determination of the $^{31}$Si half-life literature value is possible. Consequently, a considerable decrease of the measurement uncertainty is likely.

## AUTHOR INFORMATION

**Corresponding Author**

*E-mail: g.dagostino@inrim.it

## ACKNOWLEDGMENTS

This work was jointly funded by the European Metrology Research Programme (EMRP) participating countries within the European Association of National Metrology Institutes (EURAMET), the European Union, and the Italian ministry of education, university, and research (awarded project P6-2013, implementation of the new SI). The authors would like to thank the PTB workshop for cutting AVO28 crystal and Marco Santiano for the machining of the counting containers.

Table of Contents graphic

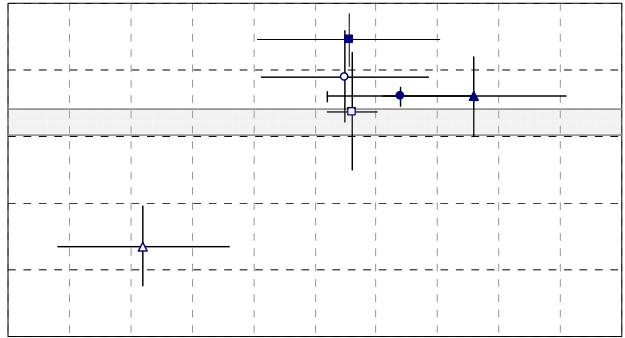